\documentclass[preprint,prd,aps,tighten,nofootinbib,amssymb]{revtex4}
\usepackage{graphicx}
\usepackage{epsfig,amssymb,amsmath}

\makeatletter
\def\Hline{%
\noalign{\ifnum0=`}\fi\hrule \@height 2pt \futurelet
\reserved@a\@xhline}
\makeatother

\newcommand{\beq}{\begin{equation}}
\newcommand{\eeq}{\end{equation}}
\newcommand{\bea}{\begin{eqnarray}}
\newcommand{\eea}{\end{eqnarray}}
\newcommand{\bear}{\begin{array}}
\newcommand {\eear}{\end{array}}
\newcommand{\bef}{\begin{figure}}
\newcommand {\eef}{\end{figure}}
\newcommand{\bec}{\begin{center}}
\newcommand {\eec}{\end{center}}
\newcommand{\non}{\nonumber}

\newcommand{\la}{\left\langle}
\newcommand{\ra}{\right\rangle}

\def\EQ#1{Eq.~(\ref{#1})}

\def\REF#1{(\ref{#1})}
\def\GEV#1{10^{#1}{\rm\,GeV}}

\def\lrf#1#2{ \left(\frac{#1}{#2}\right)}
\def\lrfp#1#2#3{ \left(\frac{#1}{#2} \right)^{#3}}
\def\oten#1{ {\mathcal O}(10^{#1})}

\newcommand{\dnf}{\Delta N_{\rm eff}}
\newcommand{\nf}{ N_{\rm eff}}
\newcommand{\mf}{m_{\rm HDM}^{\rm eff} }
\newcommand{\hdm}{{\rm HDM}}

\begin{document}
\draft
\tighten
\preprint{DESY 13-181, TU-948, ICRR-661-2013-10, IPMU13-0197}
\title{\large \bf
Axions as Hot and Cold Dark Matter
}
\author{
    Kwang Sik Jeong\,$^{a}$\footnote{email: kwangsik.jeong@desy.de},
    Masahiro Kawasaki\,$^{b,d}$\footnote{email: kawasaki@icrr.u-tokyo.ac.jp},
    Fuminobu Takahashi\,$^{c,d}$\footnote{email: fumi@tuhep.phys.tohoku.ac.jp}
    }
\affiliation{
$^a$  Deutsches Elektronen Synchrotron DESY, Notkestrasse 85,
      22607 Hamburg, Germany\\
$^b$ Institute for Cosmic Ray Research, University of Tokyo, Kashiwa 277-8582, Japan\\
$^c$ Department of Physics, Tohoku University, Sendai 980-8578, Japan\\
$^d$ Kavli IPMU, TODIAS, University of Tokyo, Kashiwa 277-8583, Japan
}

\vspace{2cm}

\begin{abstract}
The presence of a hot dark matter component has been hinted at $3 \sigma$ by a combination
of the results from different cosmological observations.
We examine a possibility that pseudo Nambu-Goldstone bosons account for both hot and cold
dark matter components.
We show that the QCD axions can do the job for the axion decay constant
$f_a \lesssim {\cal O}(10^{10})$~GeV, if they are produced by the saxion decay and
the domain wall annihilation.
We also investigate the cases of thermal QCD axions, pseudo Nambu-Goldstone bosons coupled
to the standard model sector through the Higgs portal, and axions produced by modulus
decay.
\end{abstract}

\pacs{}
\maketitle

\section{Introduction}
\label{sec:1}

The standard lambda cold dark matter (LCDM) model of cosmology provides an excellent
fit to various cosmological observations, and there is no doubt that the current
Universe is dominated by dark energy  and dark matter, while ordinary matter is only a minor
component.
Yet this apparent success does not preclude the existence of an extra component in the dark sector.

Recently it has become clear that there is a tension among different cosmological observations,
which gives a preference to a hot dark matter (HDM) component~\cite{Wyman:2013lza,Hamann:2013iba,
Battye:2013xqa,Gariazzo:2013gua}.  According to Ref.~\cite{Hamann:2013iba}, a combination
of Planck data, WMAP-9
polarization data, measurements of the BAO scale, the HST measurement of the $H_0$, Planck galaxy
cluster counts and galaxy shear data from the CFHTLens survey yields
\bea
\label{obsnf}
\Delta N_{\rm eff} &=& 0.61 \pm 0.30,\\
\label{obsmf}
\mf&=& (0.41 \pm 0.13)\, {\rm eV},
\eea
at $1\sigma$,
where $\Delta N_{\rm eff}$ denotes the additional effective neutrino species and
$\mf$ denotes the effective HDM mass
(see Eqs.~(\ref{defdnf}) and (\ref{defmf}) for the definition).
The other groups obtained similar results.

Let us focus on the extension of the LCDM cosmology by adding a HDM component, although
the above results do not exclude the existence of massless dark radiation.\footnote{
 If there are both massless dark radiation and HDM, there
will be three coincidences of the abundances of
$\rho_{\rm baryon} \sim \rho_{\rm CDM}$, $\rho_{\rm photon} \sim \rho_{\rm dark\,radiation}$,
and $\rho_{\rm neutrino} \sim \rho_\hdm$.
The solution may be the dark
parallel world with particle contents and interactions that are quite similar,
if not identical, to the standard model~\cite{Higaki:2013vuv}.
}
The important difference of HDM from massless dark radiation is that it has a small but
non-zero mass\footnote{There are numerous works on dark radiation.
See e.g. Refs.~\cite{Nakayama:2010vs, Weinberg:2013kea,Jeong:2013eza} for thermal production
and Refs.~\cite{Ichikawa:2007jv} for non-thermal  production of dark radiation.
},
which calls for some explanation. The light mass could be the result of an underlying
symmetry such as shift symmetry, gauge symmetry, or chiral symmetry~\cite{Nakayama:2010vs}.
In the case of shift symmetry, the corresponding Nambu-Goldstone (NG)
boson is expected to have a small but non-zero mass as it is widely believed that there is no
exact global symmetry~\cite{Banks:2006mm}.
The effect of mass is twofold. First, the pseudo Nambu-Goldstone (pNG) bosons behave like HDM
whose effect on the cosmological observables cannot be mimicked by massless
dark radiation~\cite{Archidiacono:2013cha}. Secondly, the mass enables the pNG bosons to oscillate
around the potential minimum, and the coherent oscillations will contribute to CDM if they are
stable in a cosmological time scale.
Then there is an interesting possibility that the pNG bosons explain both HDM and CDM,
thereby providing a unified picture of the two dark components.

One of the well-studied pNG bosons is the QCD axion, which arises in association with the spontaneous
breakdown of the Peccei-Quinn (PQ) symmetry, and its mass is assumed to come predominantly
from the QCD anomaly~\cite{Peccei:1977hh}.
Not only does the axion provide the most elegant solution to the strong CP problem, but it also
contributes to dark matter. See Refs.~\cite{QCD-axion} for the review.

The purpose of this paper is to investigate a possibility that pNG bosons, especially the QCD axions,
account for both HDM and CDM, the former of which is preferred by the recent observations.
We will also discuss whether a pNG boson coupled to the standard model (SM) through the Higgs portal
as well as axions in string theory can similarly do the job.

\section{Hot Dark Matter}
\label{sec:2}
Here let us summarize the properties of hot dark matter suggested by the
observations~\cite{Wyman:2013lza,Hamann:2013iba,Battye:2013xqa,Gariazzo:2013gua}.
The HDM component is relativistic and contributes to the
total radiation energy density $\rho_{\rm rad}$ after the electron-positron annihilation and
(much) before the photon decoupling.
It is customary to express $\rho_{\rm rad}$ in terms of the photon energy density $\rho_\gamma$
and the effective neutrino species $\nf$ as
\bea
\rho_{\rm rad} &=& \left(1+ N_{\rm eff} \frac{7}{8}
\Big(\frac{T_\nu}{T_\gamma}\Big)^4 \right) \rho_\gamma,
\label{rhorad}
\eea
where $T_\gamma$ and $T_\nu (= (4/11)^{\frac{1}{3}} T_\gamma)$ are the temperature of photons
and neutrinos, respectively.
While the effective neutrino species $\nf$ is equal to $3.046$ in the standard cosmology,
it takes a larger value in the presence of extra relativistic degrees of freedom.
The additional effective neutrino species $\dnf \equiv \nf - 3.046$ is given by
\bea
\dnf &=& \frac{\rho_\hdm}{\rho_{\nu 1}},
\label{defdnf}
\eea
where $\rho_\hdm$ is the HDM energy density, and $\rho_{\nu 1} = (7 \pi^2/120)\, T_\nu^4$ is
the energy density of a single neutrino species (e.g., $\nu_e$ + $\bar{\nu}_e$).
Note that $\dnf$ is evaluated when the HDM component is relativistic.

Following Ref.~\cite{Ade:2013zuv}, we define the effective HDM mass $\mf$ as
\bea
\mf &\equiv& m_\hdm \frac{n_\hdm}{n_\nu}
\non \\
&=& (94.1\, \Omega_\hdm h^2) {\rm\, eV},
\label{defmf}
\eea
where $m_\hdm$ is the physical HDM mass, $n_\nu = (3 \zeta(3)/2 \pi^2) T_\nu^3$
is the number density of a single neutrino species,
and $\Omega_\hdm$ is the density parameter for the HDM with $h$ being the present
Hubble parameter in the unit of $100$ km s$^{-1}$ Mpc$^{-1}$.
The second equation is derived using the density parameter for the ordinary neutrinos,
$\Omega_\nu h^2 = (\sum m_\nu)/94.1$\,eV.

If the HDM is thermally distributed, $\dnf$ and $\mf$ are given by
\bea
\label{thnf}
\dnf &=& \frac{4}{7}\, x \,g \lrfp{T_\hdm}{T_\nu}{4},\\
\label{thmf}
\mf &=& \frac{2}{3}\, y \,g \lrfp{T_\hdm}{T_\nu}{3} m_\hdm \non\\
&=& \frac{2y}{3} \lrfp{7}{4x}{ \frac{3}{4}} g^\frac{1}{4} \left(\dnf\right)^\frac{3}{4} \,m_\hdm,
\eea
with
\bea
x &=& \left\{
\bear{cl}
1 & {\rm~ for~boson} \\
7/8 &{\rm~ for~fermion}
\eear
\right.,\\
y &=& \left\{
\bear{cl}
1 & {\rm~ for~boson} \\
3/4 &{\rm~ for~fermion}
\eear
\right.,
\eea
where $T_\hdm$ is the HDM temperature, $g$  is the internal degrees of freedom:
e.g. $g=1$ for a real scalar and $g=2$ for a sterile neutrino.
The HDM component becomes non-relativistic
when $T_\hdm \sim m_\hdm$, i.e., $T_\nu \sim \mf/\dnf$.

The effect on cosmological observables is similar when the HDM is non-thermally produced
by particle decay~\cite{Hasenkamp:2012ii}. To be concrete, let us suppose that it has a
monochromatic spectrum. Then $\dnf$ and $\mf$ are written as
\bea
\dnf &=& \frac{E_\hdm\,n_\hdm}{\rho_{\nu1}},\\
\mf&=& \frac{7 \pi^4}{180 \zeta(3)} \dnf \lrf{T_\nu}{E_\hdm} m_\hdm,
\label{ntmf}
\eea
where $E_\hdm$ is the energy of the HDM particle.
The HDM component becomes non-relativistic when $E_\hdm \sim m_\hdm$, i.e.,
$T_\nu \sim \mf/\dnf$ as in the case of thermal distribution.
Note that, as we shall see in the case of axions,
$m_\hdm$ can be significantly different from $\mf$ depending on the
production process and the evolution of the Universe. For instance,
the effective mass can be of order $0.1$\,eV even for a much lighter (heavier) physical mass,
if the axions are much ``colder (hotter)" than the ambient plasma, i.e., 
$E_\hdm \ll T_\nu~ (E_\hdm \gg T_\nu)$ .

Interestingly, a combination of several different observations suggests the existence
of the HDM component in the Universe as in \REF{obsnf} and (\ref{obsmf}).
In the next three sections, we consider various scenarios to examine a possibility that
pNG bosons account for both HDM and CDM.

\section{QCD Axion Dark Matter}
\label{sec:3}

One of the well-studied pNG bosons is the QCD axion. We introduce a PQ scalar $\phi$,
which develops a non-zero vacuum expectation value (vev), leading to the
spontaneous breakdown of the U(1)$_{\rm PQ}$ symmetry:
\beq
\phi \;=\; \frac{f_a+s}{\sqrt{2}}\,e^{i \theta},
\label{PQ}
\eeq
where $f_a \equiv \sqrt{2} \la \phi \ra$ is the axion decay constant. 
Throughout this paper the radial component $s$ is called the saxion.
The axion appears as a result of the spontaneous U(1)$_{\rm PQ}$ breaking.
Since the kinetic term for $\phi$ leads to
\beq
\label{saxion-axion-action}
\partial \phi^\dag \partial \phi = \frac{1}{2} (\partial s)^2 + \frac{f_a^2}{2} (\partial \theta)^2
+ f_a s (\partial \theta)^2+\frac{s^2}{2}  (\partial \theta)^2,
\eeq
the canonically normalized axion field is $a \equiv f_a \theta$.
The axion is assumed to acquire a mass predominantly from the QCD anomaly:
\bea
m_a &\simeq& 6.0 {\rm \,eV} \lrfp{f_a/N_{\rm DW}}{\GEV{6}}{-1},
\label{ma}
\eea
where $N_{\rm DW}$ is the domain wall number. In the following we will set $N_{\rm DW} = 1$ unless 
otherwise stated.

\vspace{5mm}

In the following we consider thermal and non-thermal production of the axion HDM in turn,
and then discuss the axion CDM production by the initial misalignment mechanism and the domain wall
annihilation.

\subsection{Thermal production of axion HDM}

In the early Universe, axions are produced in thermal plasma, and they contribute to HDM.
For the decay constant $f_a \lesssim  \GEV{8}$, the axions are dominantly produced by the process
$\pi + \pi \rightarrow \pi + a$, and decouple after the QCD phase transition.
The abundance of thermal axions was evaluated in Ref.~\cite{Hannestad:2005df}.
Using the results of Ref.~\cite{Hannestad:2005df}, one can estimate $\dnf$ and $\mf$ as
\vskip 0.5cm
\begin{center}
\begin{tabular}{ l l l l l l }
\Hline
$ f_a\,$ [GeV]  & $\quad g_*(T_D) $ & $\quad \dnf \quad$ & $\quad \mf $ [eV] \\
\hline
  $ 3 \times 10^6$     &   $\quad 14.54$     & $\quad 0.382$  & \quad $0.99$     \\
  \hline
  $1 \times 10^7$    &   $\quad 16.43$     & $\quad 0.325$  &  \quad $0.26$    \\
    \hline
  $3 \times 10^7$    &   $\quad 21.10$     & $\quad 0.233$  &  \quad $0.068$    \\
\Hline
\end{tabular}
\end{center}
\vskip 0.3cm
Here $g_*(T_D)$ denotes the relativistic degrees of freedom at the decoupling temperature $T_D$,
and we have used the expression of $\dnf$~\cite{Nakayama:2010vs},
\beq
\dnf = \frac{4}{7} \lrfp{g_{*\nu}}{g_*(T_D)}{\frac{4}{3}} \leq \frac{4}{7} \simeq 0.57,
\eeq
with $g_{* \nu} = 43/4$.
Therefore, the axion decay constant in the range between $f_a = 3 \times \GEV{6}$ and $1 \times \GEV{7}$
seems consistent with the observationally inferred values (\ref{obsnf}) and (\ref{obsmf}).

We here note that the decay constant in the above range is in tension with constraints from
various astrophysical arguments. One of the most tight constraints comes from the star
cooling argument~\cite{Raffelt:1985nj,Raffelt:1994ry}.
The limits however rely on the model-dependent axion couplings with photons and electrons,
which can  be significantly suppressed in a certain set-up~\cite{Chang:1993gm}.
On the other hand, the axion couplings with nucleons are constrained
by the energy loss argument of SN1987A, leading to
$f_a \gtrsim 4 \times \GEV{8}$~\cite{Raffelt:1999tx,Raffelt:2006cw}.
However the energy loss from the supernova core due to axion emission becomes ineffective
for a sufficiently small decay constant,
leaving a narrow allowed range at $f_a = \oten{6}$\,GeV, called the hadronic axion window.
Although the above range of $f_a = 3 \times 10^6 - 1 \times \GEV{7}$ is slightly above the
hadronic axion window, it is worthwhile noting that the limits from SN1987A could contain
relatively large uncertainties originated from  the adopted assumptions and treatment
of the nuclear reaction rate and the state of the nuclear matter in the supernovae core.

The cold axions are produced by the initial misalignment mechanism.
For $f_a$ in the above range, however, the abundance of axion coherent oscillations is
too small to account for the total dark matter abundance.
Alternatively, as we shall see later in this section, a right amount of axion CDM can be
produced by the domain wall annihilation.

\subsection{Non-thermal production of axion HDM}

Here we will show that the axions produced by the saxion decay can contribute to
the HDM, and in particular, it can mimic the hot dark matter with $ \mf \sim {\cal O}(0.1)$\,eV
even for $f_a \gtrsim 4 \times \GEV{8}$ satisfying the limits from SN1987A.

The saxion is produced by coherent oscillations, and its energy density often dominates or
comes close to dominating  the Universe. For instance,  if the saxion is trapped at the origin
by thermal effects, it often drives thermal inflation~\cite{Yamamoto:1985rd,Lazarides:1985ja,
Lyth:1995hj,Lyth:1995ka,Choi:1996vz,Chun:2000jr,Kim:2008yu,Choi:2009qd}. Furthermore,
in a supersymmetric theory, the saxion is a flat direction lifted
dominantly by the supersymmetry breaking effect, and therefore it is plausible that the saxion
starts to oscillate with a large amplitude, contributing to a significant fraction of the
energy of the Universe.

The saxion decays into a pair of axions with the rate
\beq
\Gamma_{s} \;=\; \frac{c}{32 \pi} \frac{m_s^3}{f_a^2},
\eeq
where $c$ depends on the details of the saxion stabilization~\cite{Chun:1995hc,Jeong:2013axf}.
Here we take $c = 1$, which is the case for (\ref{saxion-axion-action}) where U$(1)_{\rm PQ}$
is broken mainly by $\phi$, and assume a sudden decay when the Hubble parameter equals to the
decay rate, $H=\Gamma_s$.
The saxion can decay into gluons (and gluinos) as well as into Higgs bosons
in the DFSZ axion model~\cite{Dine:1981rt,Zhitnitsky:1980tq},
but for the moment we assume that the saxion mainly decays into axions.
In the following we consider two cases, in which
$(i)$ the saxion dominates the Universe before the decay and subsequently entropy
production occurs to dilute the axion density to the observationally allowed value;
$(ii)$ the saxion energy density is subdominant at the decay.

We note that, even if the saxion dominates the Universe,
a significant fraction of the saxion coherent oscillations can evaporate into thermal
plasma through the dissipation effect, suppressing the abundance of relativistic
axions~\cite{Moroi:2013tea}.\footnote{
In Ref.~\cite{Moroi:2013tea}, the axion contribution to $\dnf$ was evaluated, but
the effect of the axion mass was neglected.
We point out that such axions naturally explain the HDM suggested by the recent observations.
}
This is an attractive possibility because the saxion can reheat the Universe without
late-time entropy production.  This scenario can be approximately modelled by our analysis
on the case $(ii)$.

First let us consider the case $(i)$, in which we assume that the Universe was once dominated
by the axion radiation and subsequently a late-time entropy occurs to dilute the axion density
to the observationally allowed value.
The effective neutrino species $\Delta N_{\rm eff}$ receives a contribution
from axions according to~\cite{Choi:1996vz,Jeong:2012np}
\beq
\Delta N_{\rm eff} \;=\; \frac{43}{7} \lrfp{43/4}{g_{*R}}{\frac{1}{3}} r,
\label{dnf1}
\eeq
with
\beq
r \;\equiv\; \lrf{\rho_a}{\rho_{r}}_R,
\eeq
where $g_{*R}$ denotes the relativistic degrees of freedom at the entropy production,
and $r$ denotes the ratio of the axion energy density $\rho_a$ to the SM radiation energy
density $\rho_r$ at the entropy production.
The subscript $R$ means that the variables are evaluated at the entropy production.
Note that $r < 1$ is required for $\dnf$ to be in the allowed range of \EQ{obsnf}.

The axion has an energy equal to $m_s/2$ at the production, and it is red-shifted
as the Universe expands. What is relevant for the observation is the effective mass $m_a^{\rm (eff)}$
defined by (cf.~\EQ{ntmf})
\beq
m_a^{\rm (eff)} \;=\;  \frac{7 \pi^4}{180 \zeta(3)} \dnf \left.\frac{T_\nu}{E_a}\right|_{\nu\,{\rm dec}} m_a,
\label{maeff}
\eeq
where $T_\nu$ is the neutrino temperature, $E_a$ the axion energy, and the subscript $\nu\,{\rm dec}$
means that the variables are evaluated at the decoupling of neutrinos.
We can evaluate $m_a^{\rm (eff)}$ as follows,
\bea
\left.\frac{T_\nu}{E_a}\right|_{\nu\, {\rm dec}}
&=&  \left.\frac{s^\frac{1}{3}}{E_a}\right|_R \left.\frac{T_\nu}{s^\frac{1}{3}}\right|_{\nu\,{\rm dec}}
\non\\
&=&  \left.\frac{T_R}{E_a}\right|_R  \times \lrfp{g_{*R}}{g_{*\nu}}{\frac{1}{3}} ,
\eea
where $s$ is the entropy density, $g_{*\nu} = 43/4$ is the relativistic degrees of freedom at
the neutrino decoupling, and $T_R$ is the temperature of the SM plasma at the entropy production.
Here $\left.T_R/E_a\right|_R$ is given by
\bea
\left.\frac{T_R}{E_a} \right|_R
&=& \lrfp{\pi^2 g_{*R}}{30}{-\frac{1}{4}} \frac{2}{m_s}
 \frac{(3 \Gamma_s^2 M_p^2)^\frac{1}{4}}{r^\frac{1}{4}}.
\eea
Substituting this result into (\ref{maeff}) leads to
\bea
m_a^{\rm (eff)} &\simeq& 0.6 {\rm \,eV}
\lrfp{\Delta N_{\rm eff}}{0.6}{\frac{3}{4}} \lrfp{m_s}{\GEV{4}}{\frac{1}{2}}
\lrfp{f_a}{\GEV{9}}{-2},
\eea
where we have used (\ref{ma}).

In the case $(ii)$, the effective neutrino species $\dnf$ is similarly given by
\beq
\Delta N_{\rm eff} \;=\; \frac{43}{7} \lrfp{43/4}{g_{*d}}{\frac{1}{3}} {\tilde r},
\label{dnf2}
\eeq
with
\bea
{\tilde r} &=& \left. \frac{\rho_s}{\rho_r} \right|_{\rm decay},
\eea
where ${\tilde r}$ denotes the ratio of the saxion energy density to the radiation energy density
evaluated at the saxion decay. The effective mass is
\bea
m_a^{\rm (eff)} &\simeq& 0.4 {\rm \,eV} \lrfp{g_{*d}}{106.75}{\frac{1}{12}}
 \lrf{\Delta N_{\rm eff}}{0.6} \lrfp{m_s}{\GEV{4}}{\frac{1}{2}}
\lrfp{f_a}{\GEV{9}}{-2},
\eea
where $g_{*d}$ counts the relativistic degrees of freedom in the thermal plasma at the
saxion decay, and we have approximated $1+{\tilde r} \simeq 1$ in the above expression.
In the numerical estimate we do not use this approximation, but the results are practically
the same.

\begin{figure}[t]
\begin{center}
\hspace{-5mm}
\begin{minipage}{8cm}
\includegraphics[width=7.6cm,clip,angle=0]{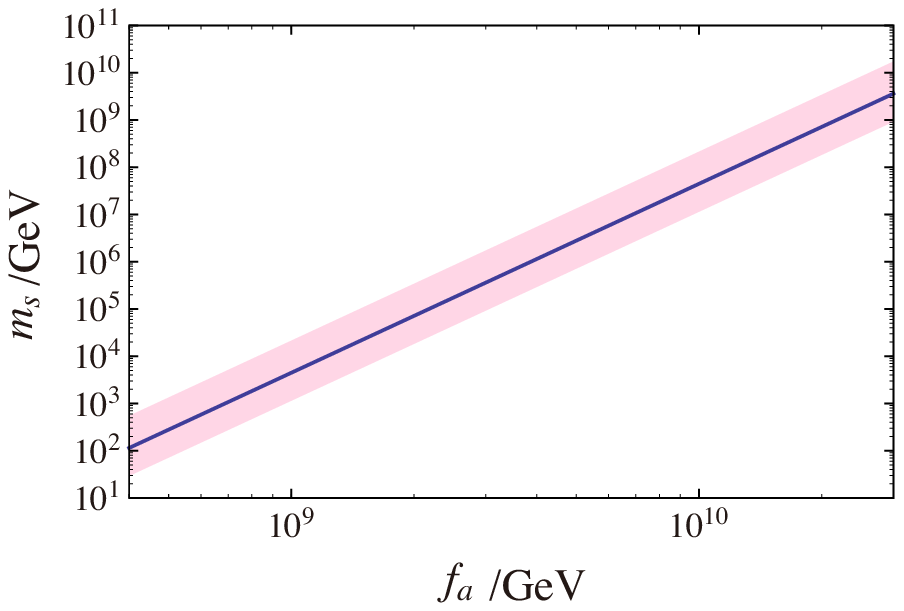}
\end{minipage}
\hspace{-5mm}
\begin{minipage}{8cm}
\includegraphics[width=7.6cm,clip,angle=0]{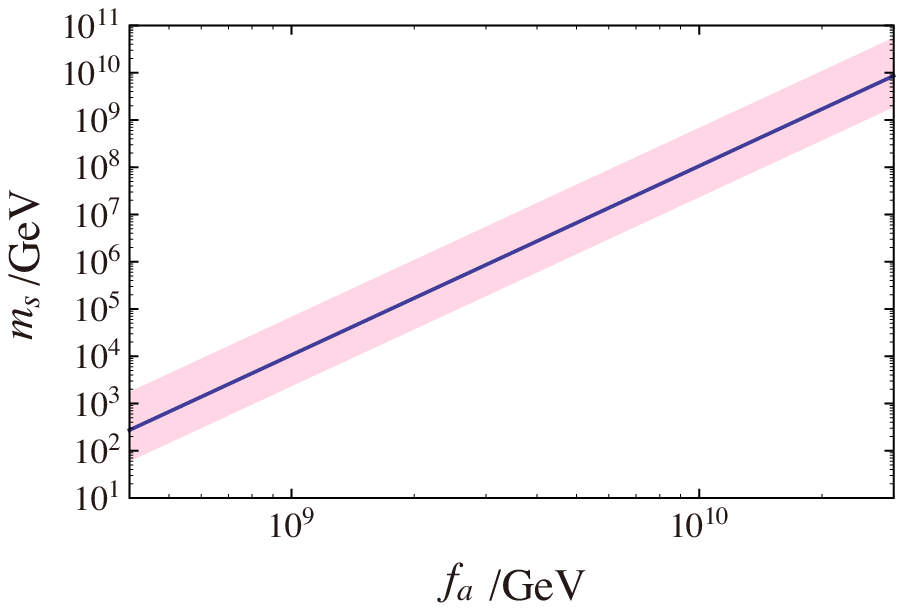}
\end{minipage}
\caption{
The HDM abundance and mass $(\dnf, m_a^{\rm eff})$ fall in the $1\sigma$ allowed values
(\ref{obsnf}) and  (\ref{obsmf}) in the shaded region on the plane of the axion decay
constant $f_a$ and the saxion mass $m_s$ for the case $(i)$ (left) and case $(ii)$ (right).
The line inside the shaded region corresponds to the center values of \EQ{obsnf} and \REF{obsmf}.
}
\label{fig:ms-fa}
\end{center}
\end{figure}

In Fig.~\ref{fig:ms-fa}, we show the $1\sigma$ allowed region for $\dnf$ and $m_a^{\rm eff}$
in the $(f_a, m_s)$ plane.
One can see that, in both cases $(i)$ and $(ii)$, the axion produced by the saxion decay can
account for the HDM for
$
m_s \approx 10^3 - 10^5 {\rm \,GeV} \,(f_a / \GEV{9})^4
$
as long as $\dnf  \sim {\cal O}(0.1)$.
Note that the axion decay constant $f_a$ is bounded above,
$f_a \lesssim  3 \times \GEV{10}$, for the perturbative stabilization of the PQ scalar.

When the saxion starts to oscillate from the origin after being trapped by thermal effects,
the saxon coherent oscillations partially evaporate to form thermal plasma through the dissipation
processes, suppressing the axion abundance~\cite{Moroi:2013tea}.
Specifically, $\dnf = {\cal O}(0.1)$ is realized for the saxion mass ranging from $\GEV{3}$
to $\GEV{4}$ at $f_a = \GEV{9}$ in a certain set-up.
Combined with the above analysis, therefore, we conclude that the axions produced from the saxion
decay naturally behave as HDM, and such axion HDM will be a natural outcome of the saxion
trapped at the origin.

\subsection{Axion CDM}

The axion coherent oscillations are produced by the initial misalignment mechanism,
and they contribute to the CDM.
Suppose that the PQ symmetry is broken during and after inflation.
Then the abundance of axion CDM is approximately given by~\cite{Bae:2008ue}
\beq
\label{omegaann}
\Omega_a h^2 \;\simeq\; 0.195\, \theta_*^2 F(\theta_*) \lrfp{f_a/N_{\rm DW}}{\GEV{12}}{1.184},
\eeq
where $\theta_* \equiv  a_*/f_a$ is the initial misalignment angle, and
$F(\theta_*)$ represents the anharmonic effect~\cite{Visinelli:2009zm},
\beq
F(\theta_*)\;=\; \left[\ln\lrf{e}{1-\frac{\theta_*^2}{\pi^2}}\right]^{1.184},
\eeq
where we have changed the exponent from the original one so as to be consistent with the axion
abundance \REF{omegaann}.
For $\theta_* = {\cal O}(1)$, the total CDM abundance can be explained by the axion coherent
oscillations with $f_a \approx \GEV{11-12}$.
Actually, however, the axion can account for the total CDM abundance even for
$f_a \lesssim \oten{10}$\,GeV, if it initially sits near the hilltop of the potential, thanks
to the anharmonic effect.
For instance, one needs to fine-tune the initial position near the hilltop at $1\,(0.01) \%$
level for $f_a \simeq 3 \times 10^{10}\,(10^{10})$\,GeV~\cite{Kobayashi:2013nva}.\footnote{
Note that the isocurvature density perturbations are enhanced toward the hilltop initial condition,
thereby tightening constraints on the inflation scale~\cite{Kobayashi:2013nva}.
}

If the PQ symmetry is restored and becomes spontaneously broken
after inflation, topological defects such as cosmic strings and domain walls are produced.
Most importantly, the axions are radiated by those topological defects.
It depends on the evolution of the string-wall network how many axions are produced.
If $N_{\rm DW}$ is equal to unity, strings and domain walls disappear soon after
the QCD phase transition due to the tension of the domain walls.
The axions radiated by the string-wall network can account for the total CDM abundance
for $f_a \approx (2.0 - 3.8) \times \GEV{10}$~\cite{Hiramatsu:2012gg}.
On the other hand, if $N_{\rm DW} \geq 2$, domain walls are stable,
leading to the cosmological domain wall problem. To avoid the cosmological catastrophe,
one needs to add small PQ symmetry breaking effect, which lifts the degeneracy among
different CP conserving vacua. As a result, the domain walls annihilate when the pressure
due to the bias becomes comparable to the wall tension~\cite{Gelmini:1988sf,Larsson:1996sp}.
According to Ref.~\cite{Hiramatsu:2012sc}, such long-lived
domain walls lead to the axion overproduction for $f_a \gtrsim
4 \times \GEV{8}$, unless the CP phase of the PQ symmetry breaking term is finely tuned
at more than $1\%$ level. Interestingly, however, the right amount of axions can be produced
in the hadronic axion window without fine-tuning of the CP phase of the PQ symmetry breaking
term.

In the case where the axion HDM is thermally produced, the required $f_a$ is of order $\GEV{6-7}$,
for which the abundance of the axion coherent oscillations is too small to account for the total DM.
As we have seen above, the right amount of axion CDM can be produced by domain wall annihilation
without fine-tuning of the CP phase of the PQ symmetry breaking operator.
On the other hand, in the case where the axion HDM is non-thermally produced by the saxion decay,
the decay constant should be in the range of $f_a = 4 \times\GEV{8} - 3 \times \GEV{10}$.
For $f_a = {\cal O}(10^{10})$\,GeV, the right amount of axion CDM can be produced by the misalignment
mechanism with a hilltop initial condition or by axion radiation from string-wall networks
with $N_{\rm DW} = 1$. For a lower $f_a$, one needs to rely on the domain wall annihilation,
which however requires a fine-tuning of the CP phase of the PQ-symmetry breaking at about $1\%$ level.
Note that the axion isocurvature perturbation is enhanced at small scales if the axions are produced
by domain wall annihilation (see footnote \ref{dw}).

\section{Nambu-Goldstone bosons through the Higgs portal}
\label{sec:4}

We consider a Higgs portal to the global U$(1)$ sector through the interaction,
\bea
\lambda |\phi|^2|H|^2,
\eea
for $\phi=(F+s)e^{ia/F}/\sqrt2$ with $F\equiv\sqrt2 \langle|\phi|\rangle$.
Here $H$ is the SM Higgs doublet developing $\langle |H^0|\rangle =v/\sqrt2$, and
$\phi$ is the scalar field which breaks spontaneously the U$(1)$ symmetry.
Then the radial scalar $s$ and the NG boson $a$ have
\bea
\label{eff-action-Higgs-portal}
{\cal L} = {\cal L}_{\rm SM} +
\mu v s h +  \frac{1}{2} \mu^\prime s^2 h
+ \frac{1}{2} m^2_s s^2 + \frac{s}{\Lambda} (\partial a)^2 + \cdots,
\eea
where $h$ is the Higgs boson with mass $m_h\simeq 125$~GeV, and the ellipsis denotes
the kinetic terms for $s$ and $a$, and also the interactions of $s$ and other hidden sector
particles if exist.
The SM and U$(1)$ sectors are connected via the $\mu$ and $\mu^\prime$ terms:
\bea
\mu = \lambda F, \quad \mu^\prime = \lambda v,
\eea
while the model-dependent parameter $\Lambda$ is generally of order $F$.
Since the radial scalar couples to $(\partial a)^2$ and mixes with the Higgs boson $h$,
integrating it out gives rise to the effective interaction
\bea
\frac{\mu}{\Lambda} \frac{m_\psi}{m^2_h m^2_s} (\partial a)^2 \bar \psi \psi,
\eea
through which the NG bosons can be thermalized with ordinary particles.
Here $\psi$ is the SM fermion with mass $m_\psi$.
The contribution of NG bosons to $\Delta N_{\rm eff}$ is not much smaller
than $4/7$ if they decouple after the QCD phase transition~\cite{Nakayama:2010vs,Weinberg:2013kea}.
For this to be the case, the radial scalar should be much lighter than the Higgs boson so
that the above interaction is strong enough for $\Lambda$ around $F$.
The NG bosons remain in thermal equilibrium until the era of muon annihilation if
the portal takes place with
\bea
\label{thermal-condition}
\frac{\mu}{\Lambda} \approx 10^{-3} \left(\frac{m_s}{200\,{\rm MeV}}\right)^2,
\eea
for $m_s$ around or above the muon mass.
One should note that the mixing between $s$ and $h$ is suppressed when
\bea
\mu \ll m^2_h/v\sim 10^2\,{\rm GeV},
\eea
implying that $\Lambda$ should be lower than about $10^5$~GeV.
In addition, $\mu/\Lambda$ is constrained to be smaller than $10^{-2}$ from the
requirement that the branching fraction of Higgs decay into NG bosons be
smaller than about 0.2 \cite{Giardino:2013bma}.
Combined with the condition (\ref{thermal-condition}), this requires the radial
scalar to be lighter than about 1~GeV.

On the other hand, if the global U(1) symmetry is only an approximate one,
the NG boson acquires a non-zero mass.
Such pNG boson may be able to account for both HDM and CDM.
We pursue this possibility in the rest of this section.

Suppose that the global U(1) symmetry is explicitly broken to the $Z_n$ subgroup
by the following interaction;
\beq
\Delta {\cal L} \;=\;  \frac{\phi^n}{ n M^{n-4}} + {\rm h.c.},
\eeq
with an integer $n \geq 5$, where $M$ is a cut-off scale.
Assuming that the above interaction does not change the potential minimum for $s$,
the potential of $a$ reads
\beq
V \;=\; \frac{m_a^2 F^2}{n^2} \left(1 - \cos\Big(\frac{n a}{F}\Big) \right),
\eeq
with the pNG boson mass given by
\beq
m_a^2 \;=\;
\frac{n}{2^{n/2-1}} \frac{F^{n-2}}{M^{n-4}}.
\eeq
For instance, the mass is about $1$ eV for the case with $F = 50$~TeV, $n=6$ and $M = M_p$:
\beq
m_a \simeq  1 {\rm \,eV}\, \lrfp{F}{50{\rm\,TeV}}{2} \lrf{M_p}{M},
\eeq
where $M_p \simeq  2.4 \times \GEV{18}$ is the reduced Planck mass.

The pNG bosons are thermalized through the Higgs portal if the radial component $s$
is relatively light.
Specifically, one can obtain $\dnf={\cal O}(0.1)$ for $m_s \approx 100$~MeV and
$F \lesssim 10^5$~GeV, while satisfying the limit coming from the invisible Higgs
decay~\cite{Weinberg:2013kea}.
The effective HDM mass is calculated as
\bea
m_a^{\rm eff} &\simeq&
0.69 \lrfp{\dnf}{0.6}{\frac{3}{4}} m_a,
\eea
by using the relation \EQ{thmf}.

One important phenomenon associated with the spontaneous break down of such discrete
symmetry is the domain wall formation.
We consider the production of pNG bosons from the domain walls in the rest of
this section, because a coherent production of the pNG bosons cannot generate the right
amount of CDM for the decay constant $F$ that leads to thermalization of pNG bosons
through the Higgs portal.

The tension of the domain wall $\sigma$ is given by
\beq
\sigma \;=\; \frac{8 m_a F^2}{n^2}.
\eeq
According to the numerical simulation~\cite{Hiramatsu:2010yu,Hiramatsu:2012sc},
the domain-wall network exhibits a scaling behavior.
Assuming the radiation dominated Universe, the scaling regime implies
\beq
\rho_{\rm dw}\;=\; 2 {\cal A} \,\sigma H,
\eeq
where $H = 1/2t$ is the Hubble parameter, and ${\cal A} \simeq 2.6$ was obtained in the
numerical simulation~\cite{Hiramatsu:2012sc}.

The domain walls should disappear before they start to dominate the Universe,
as the Universe would be significantly anisotropic. The domination takes place when
\bea
H_{\rm dom} = \frac{2{\cal A} \sigma}{ 3 M_p^2}.
\eea
Hence $H_{\rm decay} \gg H_{\rm dom}$ must be satisfied, where $H_{\rm decay}$ is the Hubble
parameter when the domain walls annihilate. In order to make the domain walls annihilate,
we need to introduce a bias that lifts the degeneracy among the $n$ vacua.
It is customary to parameterize the bias parameter as
\bea
\delta V &=& - \sqrt{2} \xi F^3 \phi e^{i \delta} + {\rm h.c.},\non\\
&=& - 2 \xi F^4 \cos{(\theta - \delta)},
\eea
where $\xi$ is a dimensionless parameter. The typical difference of the energy density between
the adjacent vacua is roughly estimated to be
\beq
\epsilon \sim \frac{8 \xi F^4}{n}
\eeq
or less.
Naively, the domain walls start to disappear when the pressure due to the bias $\epsilon$ becomes
comparable to the energy of the walls.
This happens when $\epsilon \sim \rho \simeq 2 {\cal A} \sigma H$, i.e.,
\beq
H_{\rm decay} =
\frac{1}{2 \beta {\cal A}} \frac{n \xi F^2}{  m_a},
\eeq
where we have inserted a numerical coefficient $\beta$ to represent the uncertainty of such
naive analytic estimate.
According to the numerical simulation~\cite{Hiramatsu:2010yn}, it is given by
$\beta {\cal A} \sim 18$.\footnote{
Note that this estimate based on the numerical simulation may contain a relatively large systematic
uncertainty, because it relies on extrapolating the results by many orders of magnitude.
}

There is another important parameter to evaluate the pNG boson abundance.
That is the average momentum of the pNG bosons produced by the domain wall annihilation.
It was shown that thus produced pNG bosons are marginally relativistic, and the ratio of
the averaged momentum to the mass, $\epsilon_a$, is given by
\beq
\epsilon_a \sim 1.2 - 1.5.
\eeq
Thus, the produced pNG bosons will soon become non-relativistic due to the cosmic expansion.
The precise value of $\epsilon_a$ is not important, but we will set it to be $1.5$ in the following
discussion.

The domain walls should annihilate much before the matter-radiation equality, as the
dark matter isocurvature perturbations get enhanced at small scales as
$\propto k^\frac{3}{2}$.\footnote{
This may lead to the formation of ultra-compact mini-halos. If a small fraction of
dark matter consists of thermally-produced weakly-interacting massive particles,
they may annihilate inside the mini-halos, producing an observable amount
of gamma-rays~\cite{Bringmann:2011ut}.
\label{dw}
}
In order to be consistent with the primordial density perturbations inferred from various
observations~\cite{Bird:2010mp,Hazra:2013xva,Hunt:2013bha}, we require
$H_{\rm decay} \gtrsim \oten{-22}$~eV, which corresponds to the  decay temperature
$T_d \gtrsim$ keV. The axion abundance is therefore given by
\bea
\frac{\rho_a}{s} &=& \frac{1}{\sqrt{1 + \epsilon_a^2}} \frac{2 {\cal A} \sigma H_{\rm decay} }
{\frac{2 \pi^2 g_{*s}}{45} T_d^3},
\eea
or equivalently,
\bea
\Omega_{ a} h^2
&\simeq& 0.1 \lrfp{6}{n}{2} \lrf{m_a}{1\,{\rm eV}} \lrfp{F}{2 \times 10^3 {\rm\,TeV}}{2}
\lrfp{T_d}{1\, {\rm keV}}{-1}.
\label{dwdm}
\eea
Thus, the decay constant is required to be larger than $\oten{3}$\,TeV for the pNG bosons produced
by the domain wall annihilation to comprise the total dark matter.
This is slightly too large for the pNG bosons to be thermalized through
the Higgs portal at temperature after the QCD phase transition.

The tension for obtaining both HDM and CDM can be understood as follows. In order to keep
the pNG bosons in thermal equilibrium after the QCD phase transition, its interactions should be
strong enough, placing an upper bound on $F$. On the other hand, one needs a larger value of $F$
to produce the right amount of CDM by domain walls (cf.~\EQ{dwdm}).

The crucial assumption in the above argument is that the domain wall network follows
the scaling law. We may parameterize the deviation from the scaling law as
\bea
\rho_{\rm dw} \;\approx\; \sigma H_{\rm form} \lrfp{H}{H_{\rm form}}{p}.
\eea
The scaling regime is recovered for $p=1$, and the so called frustrated domain wall network
correspond to $p=1/2$~\cite{Friedland:2002qs}. In the extreme case of the frustrated domain walls,
the domain wall abundance can be enhanced by a factor of
$T_{\rm form}/T_{\rm decay} \sim {100 {\rm MeV}}/{\rm 1 keV} \sim 10^5$.
Then we can explain the DM abundance even for $F \sim {\cal O}(10)$\,TeV, with which the thermalized
pNG bosons decouple after the QCD phase transition, leading to $\dnf = {\cal O}(0.1)$,
for a sufficiently light $m_s$.
Thus, deviation from the scaling law is required for the pNG bosons to account for
both HDM and CDM.

Alternatively, if we extend the set-up by introducing additional interactions of $\phi$,
we may be able to evade this conclusion.\footnote{
One may also consider the Higgs portal implemented by another scalar field,
for instance, by a real scalar $\varphi$.
The effective action for $\varphi$ at scales below $F$ is read off from
(\ref{eff-action-Higgs-portal}) by taking the replacement $s\to \varphi$.
For $F\gtrsim 10^3$~TeV, one can then obtain $\Delta N_{\rm eff}$ within the range
of (\ref{obsnf}) by taking $\Lambda$ smaller than $F$ and an appropriate value of $\mu$
satisfying the condition (\ref{thermal-condition}).
}
For instance,  the $\phi$ may be thermalized while it is trapped at the origin by its additional
interactions.
Then, after the phase transition, a half of the thermalized $\phi$ particles will be transformed
to the pNG bosons.
If this phase transition occurs after the QCD phase transition, $\dnf = {\cal O}(0.1)$ will be
realized.
We may also introduce multiple scalar fields by extending the global U(1) symmetry to a larger group,
which relaxes the upper bound on $F$. Also, if the mass of $\phi$ is time-dependent, it may affect
the evolution of the domain-wall network, alleviating the aforementioned tension.

If the pNG bosons are produced non-thermally by the decay of $s$, we may be able to relax the tension.
In particular, the effect of thermal evaporation may also help. We leave the detailed analysis in this
case for future work.

\section{Axions from Modulus Decay}
\label{sec:5}

Moduli fields are ubiquitous in the supergravity/string theory, and they must be
successfully stabilized in order to get a sensible low-energy theory.
Many of them can be stabilized by the flux compactification~\cite{Grana:2005jc, Blumenhagen:2006ci}
or by the KKLT mechanism~\cite{Kachru:2003aw}.
In this case the moduli fields have approximately supersymmetric spectrum, and in particular,
there are no light axions.
However, some of them may be stabilized by supersymmetry breaking effects in such a way that
their axionic fields remain light due to the shift symmetry.
The corresponding moduli fields tend to be lighter than those stabilized in a supersymmetric
fashion, and their masses are comparable to or lighter than the gravitino mass.
Such light non-supersymmetric moduli fields tend to dominate the Universe and so play an important
cosmological role. Indeed, it was recently pointed out in Ref.~\cite{Higaki:2013lra} that
axions are often overproduced by the decay of non-supersymmetric moduli, contributing to
$\dnf$. (See also Refs.~\cite{Cicoli:2012aq,Higaki:2012ar} in the context of LARGE volume scenario.)
Here we consider a case in which the produced axions have a small mass and behave
as HDM.

Let us suppose that the modulus field $\phi$ dominates the Universe and decays into
axions as well as the standard model particles.
The contribution of axions to $\dnf$ is given by~\cite{Choi:1996vz,Jeong:2012np}
\bea
\Delta N_{\rm eff}
&=&  \frac{43}{7} \lrfp{g_{* \nu}}{g_*(T_d)}{\frac{1}{3}} \frac{B_a}{1-B_a},
\eea
where $B_a$ denotes the branching fraction into axions.
The $1\sigma$ allowed range of $\dnf$ given by \EQ{obsnf} is realized
with $B_a = 0.09\pm0.04\, (0.18\pm0.06)$ for $g_* = 10.75 \,(106.75)$.
Here $T_d$ is the decay
temperature of the moduli defined by
\begin{align}
T_d \;=\; (1-B_a)^\frac{1}{4} \lrfp{\pi^2 g_*(T_d)}{90}{-\frac{1}{4}}
\sqrt{\Gamma_{\phi} M_p}\,,
\end{align}
with $\Gamma_{\phi}$ being the total decay rate of the modulus $\phi$.
Let us parameterize the total decay rate by
\beq
\Gamma_\phi \;=\; \frac{\beta}{4 \pi} \frac{m_\phi^3}{M_p^2},
\eeq
where $\beta$ is a numerical coefficient of order unity.
In order not to spoil the success of big bang nucleosynthesis, the modulus mass should
be heavier than about $100$~TeV.

The effective axion HDM mass is given by
\begin{align}
m_a^{\rm (eff)} &= \frac{7 \pi^4}{180 \zeta(3)} \dnf \left.\frac{T_\nu}{E_a}\right|_{\phi {\rm dec}}
 \lrfp{g_*(T_d)}{g_{*\nu}}{\frac{1}{3}} m_a\\
 &\simeq 0.2\, {\rm eV}\, \sqrt{\beta} \lrfp{g_{*}(T_d)}{10.75}{\frac{1}{12}}
  \lrf{\dnf}{0.6} \lrfp{m_\phi}{100{\rm\,TeV}}{\frac{1}{2}} \lrf{m_a}{1{\rm\,MeV}},
\label{maeff_moduli}
\end{align}
where we have approximated $1-B_a \simeq 1$ for simplicity.
Note that the axion mass should be of order MeV for the modulus mass $m_\phi \sim 100$\,TeV.
If the modulus field decays into the SM gauge sector, the axion HDM with such a mass
can also decay into photons, which is close to the upper limits set by the observed
$\gamma$-ray flux~\cite{Asaka:1999xd}. If the same axion constitutes CDM,
it would contribute to too much diffuse $\gamma$-ray.
If the modulus mass is heavier than $\GEV{7}$, one can avoid the observational bound as the
axion mass becomes lighter for fixed $m_a^{\rm (eff)}$.
On the other hand, if the modulus decays into the Higgs sector
through an interaction like $(\phi+ \phi^\dag)H_uH_d$ in the K\"ahler
potential~\cite{Cicoli:2012aq,Higaki:2012ar},
the axion can be stable in a cosmological time scale, and there is no such constraint
even for $m_\phi \sim 100$\,TeV.

The axion CDM can be produced by coherent oscillations. The axion CDM abundance
is given by
\beq
\Omega_a h^2 \;\simeq\; 0.3 \lrf{T_d}{10\,{\rm MeV}} \lrfp{a_*}{10^{-3} M_p}{2},
\eeq
where $a_*$ is the initial oscillation amplitude. The right amount of CDM
can be therefore produced by coherent oscillations if the initial position is sufficiently close
to the potential minimum.

In contrast to the case of the QCD axion, the physical mass of the stringy axion should be
much heavier than the effective HDM mass, which may enable the axion to decay into photons.
While one can avoid the observational limits on the axion decay, it is interesting that
$\gamma$-ray or $X$-ray can be a probe of such axion HDM/CDM.

\section{Conclusions}
\label{sec:6}

We have examined a possibility that the pNG bosons, especially the QCD axions,
account for both HDM and CDM in the Universe, the former of which has been suggested
by the recent observations (cf.~Eqs.~(\ref{obsnf}) and (\ref{obsmf})).
We divide the production process of the axion HDM into thermal and non-thermal ones.
In the thermal case, the QCD axion can explain HDM for the decay constant
$f_a \approx 3 \times 10^6 - 10^7$~GeV, which however is in tension with the SN1987A
limit even for the hadronic axion models.
On the other hand, the axion HDM can be naturally produced by the saxion decay. This is possible
for the saxion mass ranging from $\oten{3}$~GeV to $\oten{10}$~GeV and $f_a \lesssim 3 \times
\GEV{10}$.

Note that the non-thermally produced axions need to be ``colder" than the ambient plasma,
in order to explain the hierarchy between the effective HDM mass of ${\cal O}(0.1)$~eV  and the
physical axion mass $m_a = 0.006 $\,eV$(f_a/\GEV{9})^{-1}$ (cf.  \EQ{ntmf}).
We have discussed two cases in which such axions are produced.
In the case $(i)$, there is a late-time entropy production which dilutes the axions produced
by the saxion decay, assuming that the saxion dominates the Universe and decays dominantly
into a pair of axions.
In the case $(ii)$, the saxion decays into a pair of axions when it is subdominant.
Our analysis can be also applied to the case where the saxion coherent oscillations
partially evaporates into plasma after being trapped at the origin by the thermal
effects~\cite{Nakayama:2010vs}.
We have pointed out that the axion HDM can be a natural outcome of the saxion trapped
at the origin.
The axion CDM can be produced by either the initial misalignment mechanism or domain wall
annihilation.

We have also discussed the pNG bosons coupled through the Higgs portal.
While the domain walls associated with the spontaneous U(1) breaking can
be the source of the pNG CDM, the required decay constant $F \gtrsim 10^3$~TeV
is too large to keep pNG bosons in thermal equilibrium after the QCD phase transition.
Therefore it is difficult to explain both HDM and CDM simultaneously with the pNG boson
coupled to the SM through the Higgs portal, and some extension of the set-up
or deviation from the scaling law of the domain-wall network is required.

Finally we have studied the axions produced by modulus decay, which is considered to
take place generically~\cite{Higaki:2013lra}. Such axions can behave like HDM for the
axion mass of ${\cal O}(1)$\,MeV for the modulus mass $m_\phi \sim 100$\,TeV. In contrast
to the case of the QCD axion, the produced axions are more energetic than the ambient 
plasma. The right abundance of axion CDM can be generated by the coherent oscillations.
Interestingly, the X-ray or gamma-ray can be a probe of such axion dark matter as well as
the coupling of the corresponding modulus to the SM sector.

\section*{Acknowledgment}
FT thanks Kazunori Nakayama for discussion on the thermal dissipation effect of 
the QCD saxion. 
This work was supported by Scientific Research on Innovative Areas (No.~24111702 [FT], No.~21111006 [MK, FT],
and No.~23104008 [FT]), Scientific Research (A) (No.~22244030 and No.~21244033) [FT],
Scientific Research (C) (No.~25400248) [MT],
JSPS Grant-in-Aid for Young Scientists (B) (No.~24740135) [FT], and Inoue Foundation for Science [FT].
This work was also supported by World Premier International Center Initiative
(WPI Program), MEXT, Japan [MK, FT].

\end{document}